\documentclass[12pt,preprint]{aastex}

\usepackage{amsmath}
\usepackage{rotating}		
\usepackage{morefloats}
\usepackage{graphics}
\usepackage{epsfig}
\usepackage{natbib}
\shorttitle{Far-UV Continuum in Disks II: CO}
\shortauthors{Schindhelm et al.}

\begin{document}

\title{Characterizing CO Fourth Positive Emission in Young Circumstellar Disks}

\author{
Eric Schindhelm\altaffilmark{1}, Kevin France\altaffilmark{1},  Eric B. Burgh\altaffilmark{1},
Gregory J. Herczeg\altaffilmark{2},  James C. Green\altaffilmark{1}, Alexander Brown\altaffilmark{1}, Joanna M. Brown\altaffilmark{3}, Jeff A. Valenti\altaffilmark{4}}

\altaffiltext{1}{Center for Astrophysics and Space Astronomy, University of Colorado, 389 UCB, 
Boulder, CO 80309, USA}
\altaffiltext{2}{Max-Planck-Institut f\"{u}r extraterrestriche Physik, Postfach 1312, 85741 Garching, Germany}
\altaffiltext{3}{Harvard-Smithsonian Center for Astrophysics, 60 Garden Street, Cambridge, MA 02138, USA}
\altaffiltext{4}{Space Telescope Science Institute, 3700 San Martin Drive, Baltimore, MD 21218, USA}

\received{}
\revised{}
\accepted{}
\ccc{}
\cpright{AAS}{}

\begin{abstract}
Carbon Monoxide is a commonly used IR/sub-mm tracer of gas in protoplanetary disks.  We present an analysis of ultraviolet CO emission in \textit{HST}-COS spectra for 12 Classical T Tauri stars.  Several ro-vibrational bands of the CO $A^1\Pi$ - $X^1\Sigma^+$ (Fourth Positive) electronic transition system are spectrally resolved from emission of other atoms and H$_2$.  The CO $A^1\Pi$ $v'$=14 state is populated by absorption of Ly$\alpha$ photons, created at the accretion column on the stellar surface.  For targets with strong CO emission, we model the Ly$\alpha$ radiation field as an input for a simple fluorescence model to estimate CO rotational excitation temperatures and column densities.  Typical column densities range from $N_{CO}$ = 10$^{18}$ - 10$^{19}$ cm$^{-2}$.  Our measured excitation temperatures are mostly below $T_{CO}$ = 600 K, cooler than typical M-band CO emission.  These temperatures and the emission line widths imply that the UV emission originates in a different population of CO than that which is IR-emitting.  We also find a significant correlation between CO emission and the disk accretion rate $\dot M_{acc}$ and age.  Our analysis shows that ultraviolet CO emission can be a useful diagnostic of CTTS disk gas.

\end{abstract}


\clearpage

\section{Introduction}

Classical T Tauri stars (CTTSs) provide the opportunity to witness the birth of extra-solar planetary systems.  Their disks of gas and dust are the environments in which Jovian planets form \citep{Dullemond2010}, a process that we are only beginning to understand.  From the ground, carbon monoxide (CO) is the easiest molecule to observe in emission from these disks, and has been studied from near-IR to mm wavelengths \citep{Salyk2008,Dutrey2008,Sturm2010}.  Such observations sample regions from $r_{in} <$ 1 AU \citep{Najita2003} to $r_{out} >$ 800 AU \citep{Pietu2007}.  The second most abundant molecule in the galaxy (after H$_2$), CO plays a significant role in the formation of planetary systems.  As H$_2$ does not emit strongly in the IR, CO is an important tracer of disk gas.  CO abundance varies greatly in different environments throughout the Milky Way.  In the diffuse ISM, the CO/H$_2$ ratio ranges from 10$^{-7}$ to 10$^{-6}$ \citep{Burgh2007}; in dense clouds, it can be 10$^{-4}$ \citep{Lacy1994}, while comets in the solar system exhibit a ratio of $\sim$10 \citep{Lupu2007,Feldman2009}.  During the process of star and planet formation, the CO and H$_2$ may be selectively removed from or added to the gas phase in a way that changes the CO/H$_2$ ratio \citep{VMadjar1994,France2011b}.  By probing CO in protoplanetary disks, we can constrain models of planetary system formation.

Electronic transitions of CO have recently been identified in the $Hubble$ $Space$ $Telescope$ Cosmic Origins Spectrograph (\textit{HST}-COS) spectra of several CTTSs \citep{France2011b}.  Several vibrational bands of the $A^1\Pi$ - $X^1\Sigma^+$ (Fourth Positive) system were measured in emission for V4046 Sgr and RECX 11, and in absorption in HN Tau.  Lyman $\alpha$ (Ly$\alpha$) photons produced in the accretion funnel flow excite the CO, populating its $A^1\Pi$ $v'$ = 14 state; several of the resulting (14 - 2) through (14 - 12) transitions are detected between 1280 \AA\ and 1720 \AA.  Emission from the (0 - 1) 1600 \AA\ and (0 - 3) 1713 \AA\ bands also appears, potentially due to C IV photoexcitation of CO.  The shape and strength of these emission bands vary with temperature ($T_{CO}$) and column density ($N_{CO}$) such that these parameters can be determined.  Circumstellar CO can be observed in absorption as well; a collisionally-excited distribution of gas at $T_{CO} = 500 \pm 200$ K is able to reproduce several of the ($v'$ - 0) absorption bands in the HN Tau spectrum for log$_{10}(N_{CO})$ = 17.3 $\pm$ 0.3 cm$^{-2}$.  France et al. (2011, in press) also find CO in absorption in AA Tau, with log$_{10}(N_{CO})$ = 17.5 $\pm$ 0.5 cm$^{-2}$.  

We extend this work by conducting a survey of COS spectra for 12 CTTSs in greater detail.  Our aim in this paper is to characterize $T_{CO}$ and $N_{CO}$ in a variety of CTTSs, so as to understand correlations between the UV-emitting CO (UV-CO) and the disk properties that characterize these systems.  The UV-CO can also be compared with CO previously detected in these disks at IR and mm wavelengths.  We model the Ly$\alpha$ radiation field assuming absorption from an outflow and interstellar H I, using this as an input for simple fluorescence models.   These models enable us to estimate $T_{CO}$ and $N_{CO}$ for targets with high signal-to-noise (S/N) CO emission.  In \S\ref{obs} we detail the COS observations, then discuss the CO emission in section \S\ref{coemission}.  In \S\ref{analysis} we summarize our model for the Ly$\alpha$ profiles as well as the CO fluorescence model.  In \S\ref{results} we discuss the results of this analysis, and our ability to constrain the temperatures and column densities of CO in each of the targets. We also compare these empirically derived properties with other previously measured parameters from the literature.

\section{Observations} 
\label{obs}

We investigated a sample of 12 CTTSs observed by $HST$-COS; the observational parameters are listed in Table \ref{TBL:Observing_Log} in order of observation date.  These were the first CTTSs observed with COS.  Spectra from several of these targets have already been presented in the literature.  \citet{Ingleby2011} studied accretion rates in RECX 11 and RECX 1, and \citet{Yang2011} demonstrated Ly$\alpha$ photon absorption and re-emission by H$_2$ in V4046 Sgr and DF Tau.  \citet{France2011a} characterized the continuum emission of V4046 Sgr and DF Tau, while \citet{France2011b} identified CO Fourth Positive emission and absorption in the spectra of V4046 Sgr, RECX 11, and HN Tau.  This paper builds upon the work of \citet{France2011b}, advancing the models for analyzing the CO emission of a survey of CTTSs.

Each spectrum consists of 2 - 4 orbits with the G130M and G160M gratings \citep{Osterman2011}, yielding a moderate resolving power of $\lambda/\delta\lambda \sim$18,000.    As there is no field stop in the instrument, the entire objects were sampled without appreciable spatial resolution relative to the size of the disks.  We used CALCOS, the COS calibration pipeline, to reduce the data and then created coadded spectra from 1140 \AA\ to 1760 \AA.  The data reduction pipeline creates inaccurately small error bars in low count regions of the spectrum \citep{Froning2011}.  This gives excessive weight to continuum points in the least squares fits of the emission feature.  We empirically determined that this was generally the case for data points with errors more than three times smaller than the mean 1$\sigma$ continuum error.  Any errors smaller than this limit were thus set to the limit value.  

Most targets belong to the  $\eta$ Chamaeleontis and Taurus-Auriga star formation regions, with properties described in Appendix \ref{targets}.  Table \ref{Target_Parameters} lists a variety of target parameters relevant to this work, including visual extinction (A$_V$), accretion rate  ($\dot M_{acc}$), inclination, and age.

\section{Ly$\alpha$-excited CO Emission}
\label{coemission}

CO UV emission appears in 7 of the 12 targets in our survey.  We find strong CO (14 - $v''$) emission in four targets (V4046 Sgr, DF Tau, RECX 15, and DM Tau), and weaker but identifiable emission in RECX 11, HD135344 B, and DE Tau.  The CO emission bands seen are listed in Table \ref{TBL:Band_Ratios}.  The (14 - 1), (14 - 6), (14 - 9), and (14 - 11) bands are unidentifiable due to either low branching ratios or overlying emission from other molecules and ions.  We list the branching ratios $B$($v'$ - $v''$) for each band, calculated by summing the $A$ values of each $J$ transition in a given band, and dividing by the sum of all ($v'$ - $v''$) transitions for a given $v'$.  Despite the relatively low $A$ values of the overlying (14 - 0) transitions (1.32$\times$10$^4$ s$^{-1}$ for $J''$ = 0), there is sufficient Ly$\alpha$ flux to yield strong CO emission.   C IV-pumped emission bands are also seen in several targets; we present this evidence in \S\ref{civfluxes}.  The Ly$\alpha$-excited emission is the primary subject of our analysis.

\begin{figure}
  \centering
  \includegraphics[width=16.5cm]{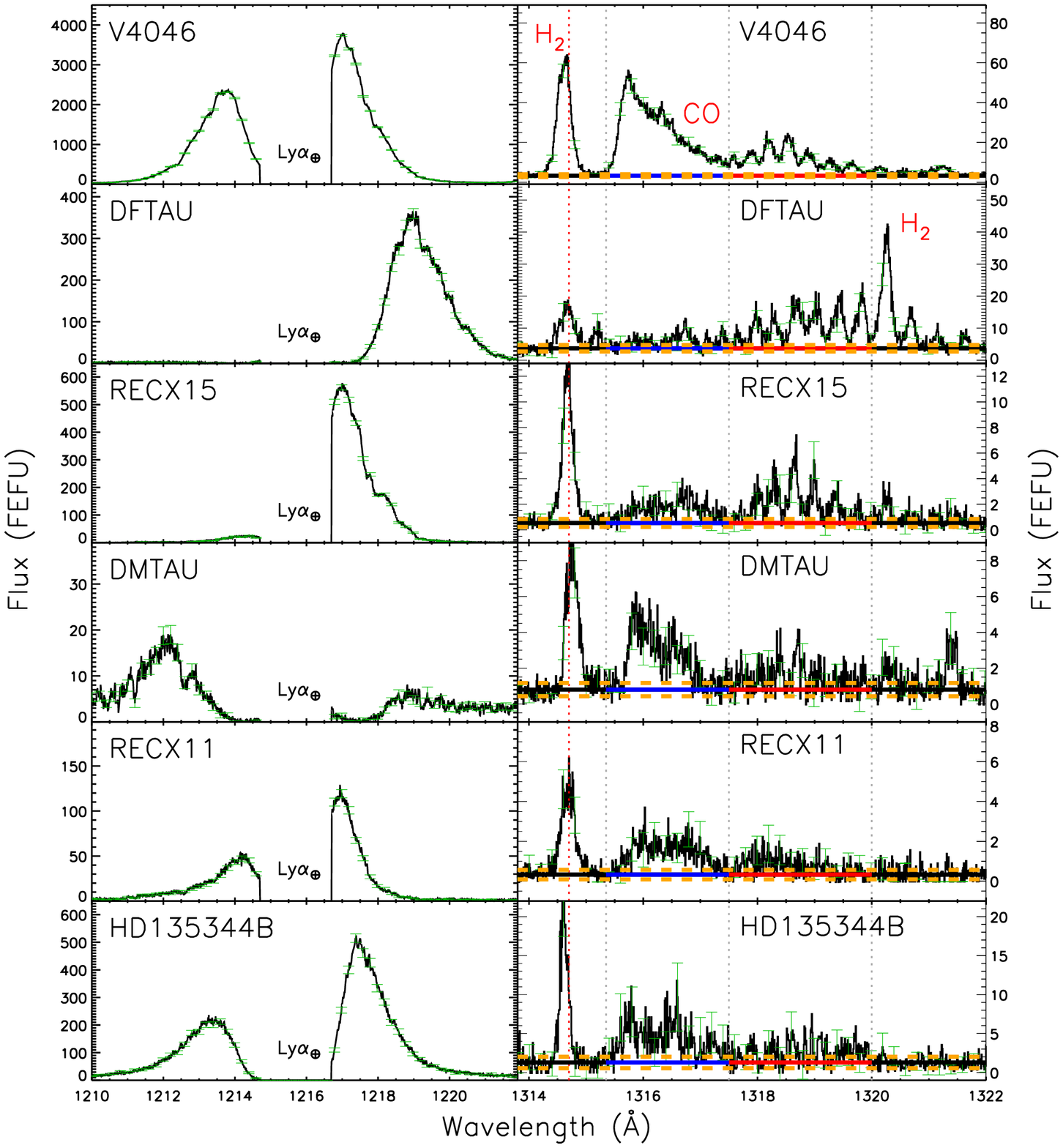}
  \caption{Extinction-corrected Ly$\alpha$ profiles (left column) with geocoronal Ly$\alpha$ removed, and CO (14 - 3) emission bands (right) for each of the sample targets.  In the CO bands, an H$_2$ line is identified with a vertical dotted red line, and the dotted grey lines bound the high and low-$J$ sides of the (14 - 3) band.  The horizontal black/blue/red line identifies the local continuum, with the dashed orange lines at the 1$\sigma$ error in the continuum average.  Representative error bars are shown in green.}
   \label{fig:Lyman_1315}
  \end{figure}

\begin{figure}
  \centering
  \includegraphics[width=16.5cm]{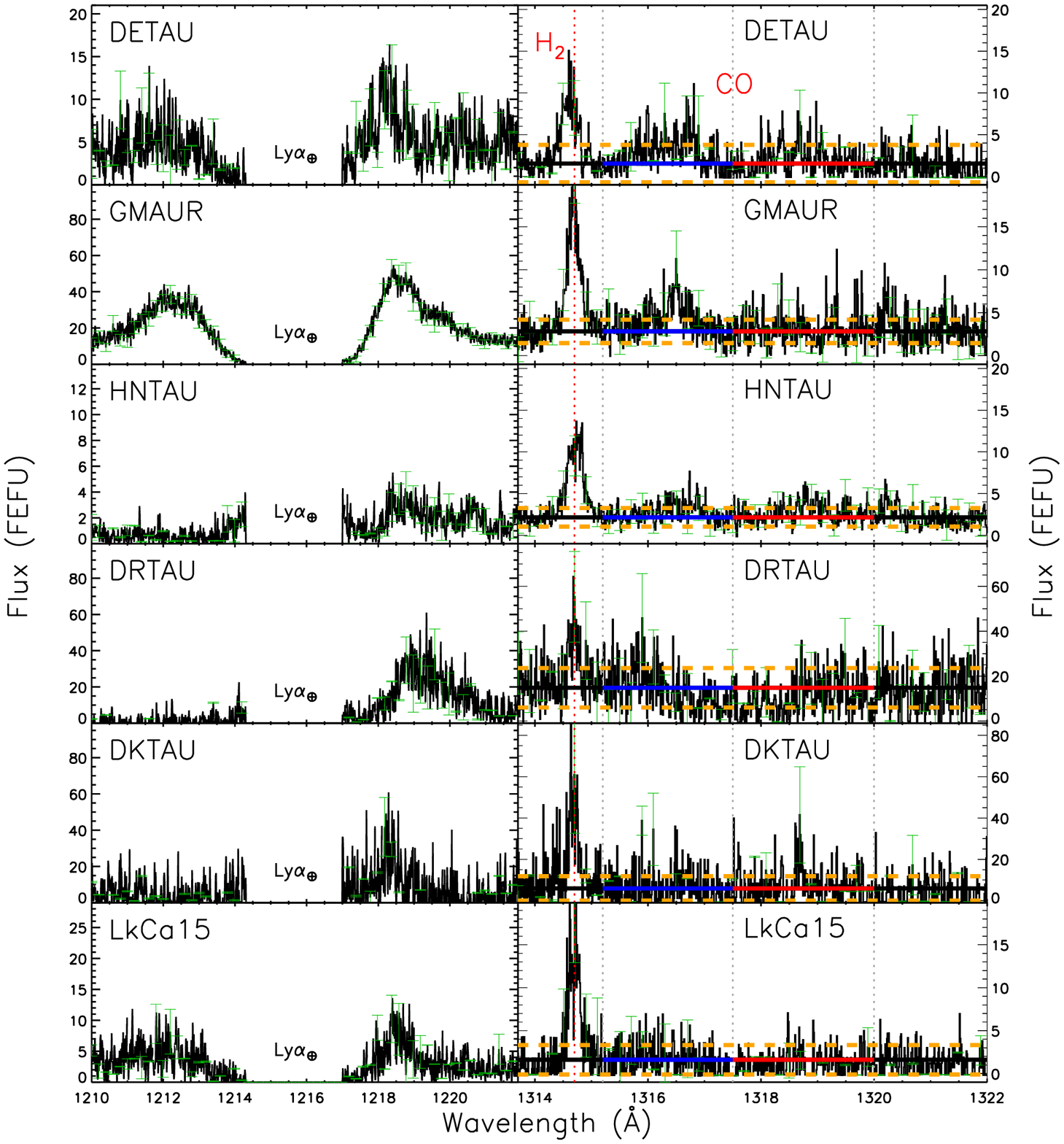}
  \caption{Extinction-corrected Ly$\alpha$ profiles (left column) with geocoronal Ly$\alpha$ removed, and CO (14 - 3) emission bands (right) for each of the sample targets.  In CO bands, an H$_2$ line is identified with a vertical dotted red line, and the dotted grey lines bound the high and low-$J$ sides of the (14 - 3) band.  The horizontal black/blue/red line identifies the local continuum, with the dashed orange lines at the 1$\sigma$ error in the continuum average. Representative error bars are shown in green.  DR Tau and LkCa 15 are non-detections.}
   \label{fig:Lyman_1315_non}
  \end{figure}

Figures \ref{fig:Lyman_1315} and \ref{fig:Lyman_1315_non} show the observed Ly$\alpha$ profiles and (14 - 3) CO bands for the sample. In this paper we use femto-erg flux units (FEFU, where 1 FEFU = 1$\times$10$^{-15}$ ergs cm$^{-2}$ s$^{-1}$ \AA$^{-1}$) as a standard flux unit.  The geocoronal Ly$\alpha$ emission in the middle of the Ly$\alpha$ profile is set to zero to avoid confusion with target Ly$\alpha$ emission.  For future discussion, we refer to the Ly$\alpha$ emission on the longer wavelength side of the interstellar absorption trough as the ``red wing", and the emission on the shorter wavelength side the ``blue wing".  The CO plots in these figures also separate the CO bands into a ``high-$J$'' and ``low-$J$'' side, identified by the color of the continuum.  The blueward CO emission comes from the lower-$J$ transitions (0 - 15), which are closely spaced relative to the instrument resolution, forming a ``bandhead''.  The redward CO emission is composed of higher $J$ transitions, which are well separated in several targets.  There is a general correlation between the relative amount of flux in the Ly$\alpha$ blue and red wing and the relative amount of flux seen in the low-$J$ and high-$J$ CO emission.  For example, in RECX 15 and DF Tau, weak blue wing Ly$\alpha$ emission correlates with the lack of low-$J$ emission in their CO bands, whereas in V4046 Sgr we observe strong blue and red emission in both the Ly$\alpha$ and CO profiles.  This will motivate our modeling, as discussed in \S\ref{analysis}.

\begin{figure}
  \centering
  \includegraphics[width=16.5cm]{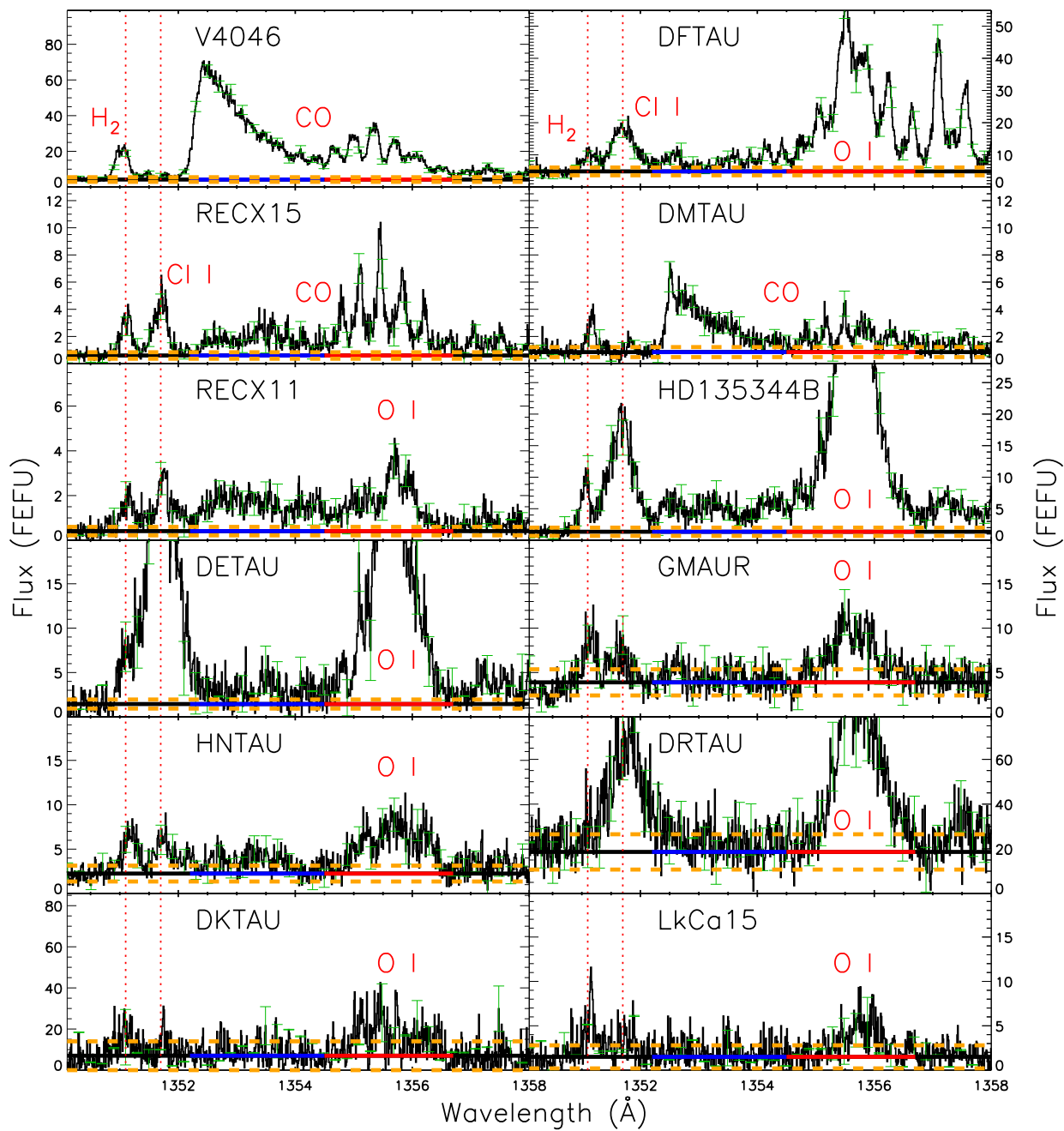}
  \caption{(14 - 4) transitions for targets as identified.  The horizontal lines show continuum levels, with dashed orange lines indicating the upper and lower limits for summation.  The solid blue and red lines mark the wavelength range for the high and low-$J$ regions as described in the text; vertical gray lines bound these regions.  Vertical dashed red lines mark contaminating lines.  Representative error bars are shown in green.  H$_2$ and Cl I emission lines are identified with red-dotted lines, and O I emission features are identified in the spectra in which they appear.}
   \label{fig:1352}
  \end{figure}

The (14 - 3) band is mostly free of emission from other molecules or atoms; only the weak H$_2$ (4 - 2) P(19) line at 1316.55 \AA\ appears in all targets.  There is an H$_2$ (3 - 2) P(18) line at 1320.22 \AA, but it appears strongly only in DF Tau.  Overall, the low-$J$ bandhead and separated high-$J$ lines of the (14 - 3) band are uncontaminated, which is crucial for determining $N_{CO}$ and $T_{CO}$.  For this reason, we primarily use the (14 - 3) band for analysis in all targets.  By branching ratios, the (14 - 4) band is stronger, however strong OI ] (1355.71 \AA) emission lines appear over the high-$J$ portion of the band in most targets.  Figure \ref{fig:1352} shows the (14 - 4) band for our sample, with similar continuum denotations as the (14 - 3) band.  Only V4046 Sgr, RECX 15, and DM Tau are free from contamination on the high-$J$ side.  For these three targets we analyze both the (14 - 3) and (14 - 4) bands.

In V4046 Sgr, DF Tau, DM Tau, and RECX 15, the high-$J$ lines are well separated.  The high-$J$ lines of DF Tau, RECX 15, and DM Tau appear at the width of the COS line-spread-function.  In RECX 11 and HD 135344B the CO high-$J$ lines are not distinguishable from one another, although whether this is physical (due to velocity broadening of close-in CO) or merely due to the lower S/N in the band is not clear.   The Ly$\alpha$ profile of DE Tau is highly absorbed, yet the (14 - 3) features are evident to the eye.  HN Tau, DR Tau, DK Tau, LkCa 15, and GM Aur show little CO emission above the continuum.  Of those, GM Aur is the only target to have a Ly$\alpha$ profile that is mostly unabsorbed by circumstellar and interstellar H I.

\section{Method and Modeling Results}
\label{analysis}

\subsection{Ly$\alpha$ and CO Fluorescence Model}

\begin{figure}
  \centering
  \includegraphics[width=16cm]{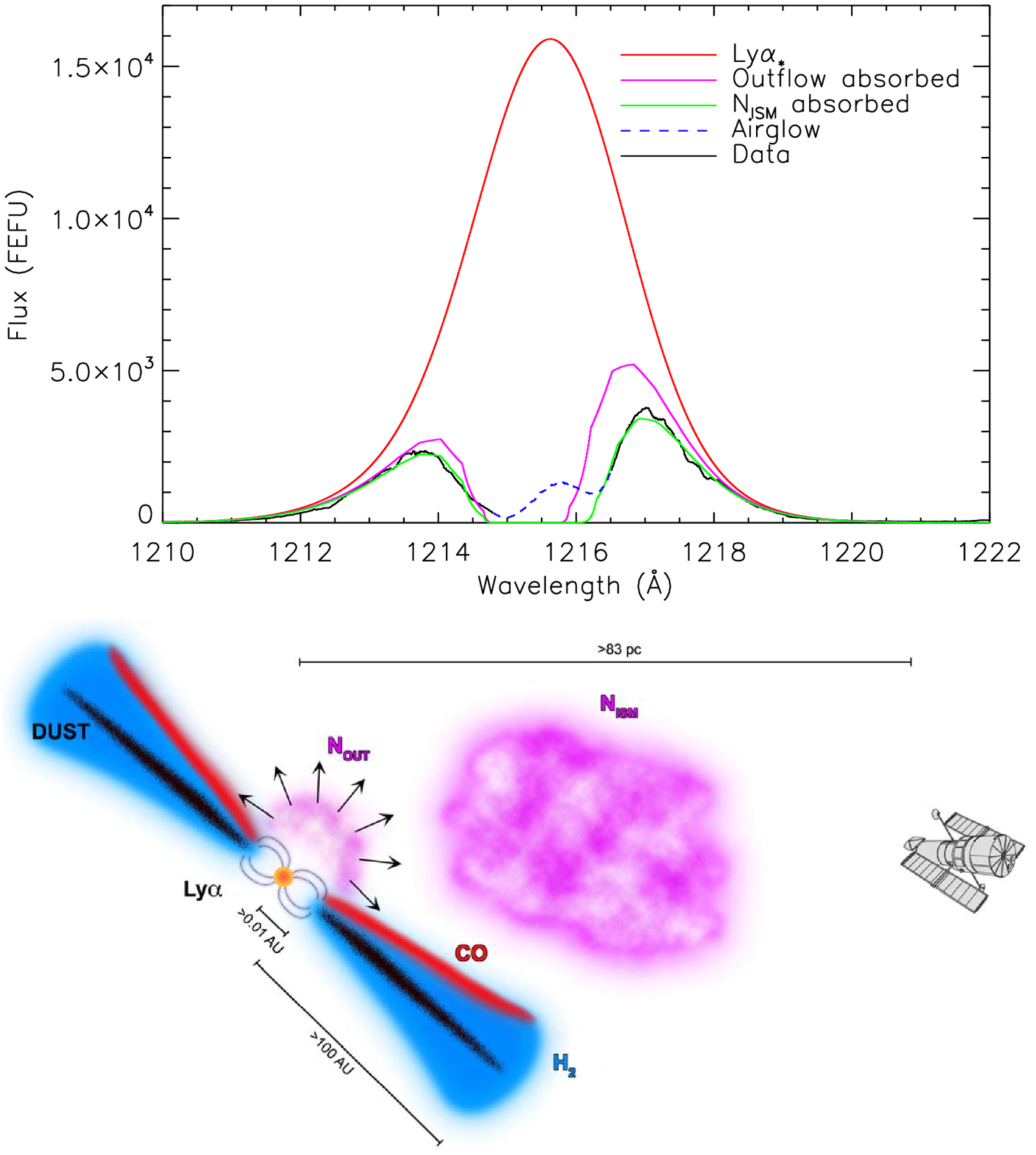}
  \caption{V4046 Sgr Ly$\alpha$ data with overlaid model profiles (top), and drawing of disk that produces the flux we see (bottom).  Ly$\alpha$ emission (red profile) is created near the star, then absorbed by an outflow (magenta profile), and additionally absorbed by interstellar H I (green profile) to produce the observed data.  The CO in the disk sees the outflow-absorbed emission.  Disk height is not to scale relative to extent.}
   \label{fig:Lyman_description}
  \end{figure}

\citet{France2011b} utilized a simple fluoresence model to calculate the temperature $T_{CO}$ and column density $N_{CO}$ for two CTTSs.  However, they used the observed Ly$\alpha$ emission to excite the CO.   In our full sample, the targets demonstrate the need for a more detailed estimation of the incident Ly$\alpha$ radiation; using the observed Ly$\alpha$ profiles as the incident radiation field cannot reproduce the observed CO band shape.  For example, with DF Tau, more red wing Ly$\alpha$ emission than is seen is necessary to yield the strong CO (14 - 3) high-$J$ line emission observed, relative to its weak low-$J$ emission.  We therefore introduce a straight-forward model of the physical situation, shown in Figure \ref{fig:Lyman_description}, which we use to re-construct the incident radiation.  Ly$\alpha$ emission is created at the accretion funnel flow near the proto-stellar surface, as well as in the photosphere itself.  The plot in Figure \ref{fig:Lyman_description} shows such a model Ly$\alpha$ profile for V4046 Sgr in red.  Along our line of sight the Ly$\alpha$ is then absorbed by H I in an outflow (magenta), followed by interstellar gas (green).  Our method consists of two steps: first, we create a series of model Ly$\alpha$ profiles that fit the observed profile after absorption by outflow and ISM H I.  Second, we use each outflow-absorbed model Ly$\alpha$ emission in a CO fluorescence model to match the observed CO emission.  

The observed Ly$\alpha$ emission is modeled as:

\begin{equation}
I_{obs} = C_{Ly\alpha}+I_{Ly\alpha}e^{-(\tau_{out}+\tau_{ISM})}
\end{equation}

\noindent where C$_{Ly\alpha}$ is the continuum intensity near Ly$\alpha$ and the source Ly$\alpha$ profile $I_{Ly\alpha}$ is approximated as the sum of two Gaussian emission components, parameterized by peaks $I_1$ and $I_2$, as well as widths $\sigma_1$ and $\sigma_2$.  The first Gaussian is at rest relative to the star, while the second Gaussian component is centered at a velocity $v_2$.  The outflow absorber $\tau_{out}$ is characterized by a Voigt profile with $N_{out}$ at a velocity $v_{out}$, and is necessary to match blue-shifted absorption seen in most of the targets.  The ISM absorber $\tau_{ISM}$ is determined by a Voigt profile with $N_{out}$ at the local standard of rest (LSR).  A$_V$ values from Table \ref{Target_Parameters} and R$_V$ = 4.0 were used to de-redden the data, according to the extinction curve from \citet{Cardelli1989}.  Seven parameters determine the two emission (stellar and accretion shock) and two absorption (interstellar and outflow) components.  As a result, many parameter combinations exist that in the end fit the observed Ly$\alpha$ profile.  However, only some will adequately produce the observed CO emission, once run through the fluorescence model.

Our CO fluorescence model uses a method similar to that implemented by \citet{France2011b}.  The ground rotational levels are populated thermally with $T_{CO}$, using Dunham coefficients from \citet{George1994} to calculate ground state energy values.   The Ly$\alpha$ flux absorbed by the (14 - 0) band is re-emitted according to the branching ratios for the (14 - $v''$) transitions.  However, instead of using the observed Ly$\alpha$ flux, we excite the CO with the outflow-absorbed profile.  For each Ly$\alpha$ profile produced in the first step, we produce a model CO emission band to fit to the data.  The variable parameters in the CO model are the gas rotational temperature ($T_{CO}$), column density ($N_{CO}$), and the velocity broadening of the emitting gas ($\Delta v_{broad}$).  This $\Delta v_{broad}$ value is the full width at half maximum (FWHM) of the Gaussian emission line, assuming the broadening is primarily due to orbital motion.  For a continuum, we use the average continuum level from nearby areas of the spectrum.  Assuming that the electronic ground state population can be described by a thermalized population, the resulting CO emission is primarily dependent on $N_{CO}$ and $T_{CO}$.  As in \citet{France2011b}, we make no assumptions about the spatial distribution of the CO around the star.  Due to the low transition probabilities of the (14 - 0) band ($\sim$ 10$^4$ s$^{-1}$), optical depth effects are minimal; therefore, $N_{CO}$ and the filling fraction of the CO are degenerate.

There are limitations to our assumption of a single CO rotational temperature, as multiple temperatures are possible throughout the disk. The gas disk temperature varies based on height in the disk and distance from the star. As the entire disk is sampled in the COS aperture, CO may be observed at multiple temperatures. However, a single temperature model does enable us to determine general characteristics of the CO in a large sample with minimal parameters. Future models of individual targets should take into account the effects of density and temperature on the radiative transfer of the Ly$\alpha$ radiation and fluorescent emission.

\subsection{Results}
\label{results}

In half of the positive detection targets, the uncertainties on $N_{CO}$ are less than 0.5 dex and the uncertainties on $T_{CO}$ are less than  $\sim$200 K.  High S/N in both the Ly$\alpha$ and CO profile are required for this.  In most of the positive detections, values for log$_{10}$($N_{CO}$) are between 18 and 19 and temperatures lie between 100 and 600 K.

\subsubsection{Ly$\alpha$ Model Fits}

\begin{figure}
  \centering
  \includegraphics[width=15cm]{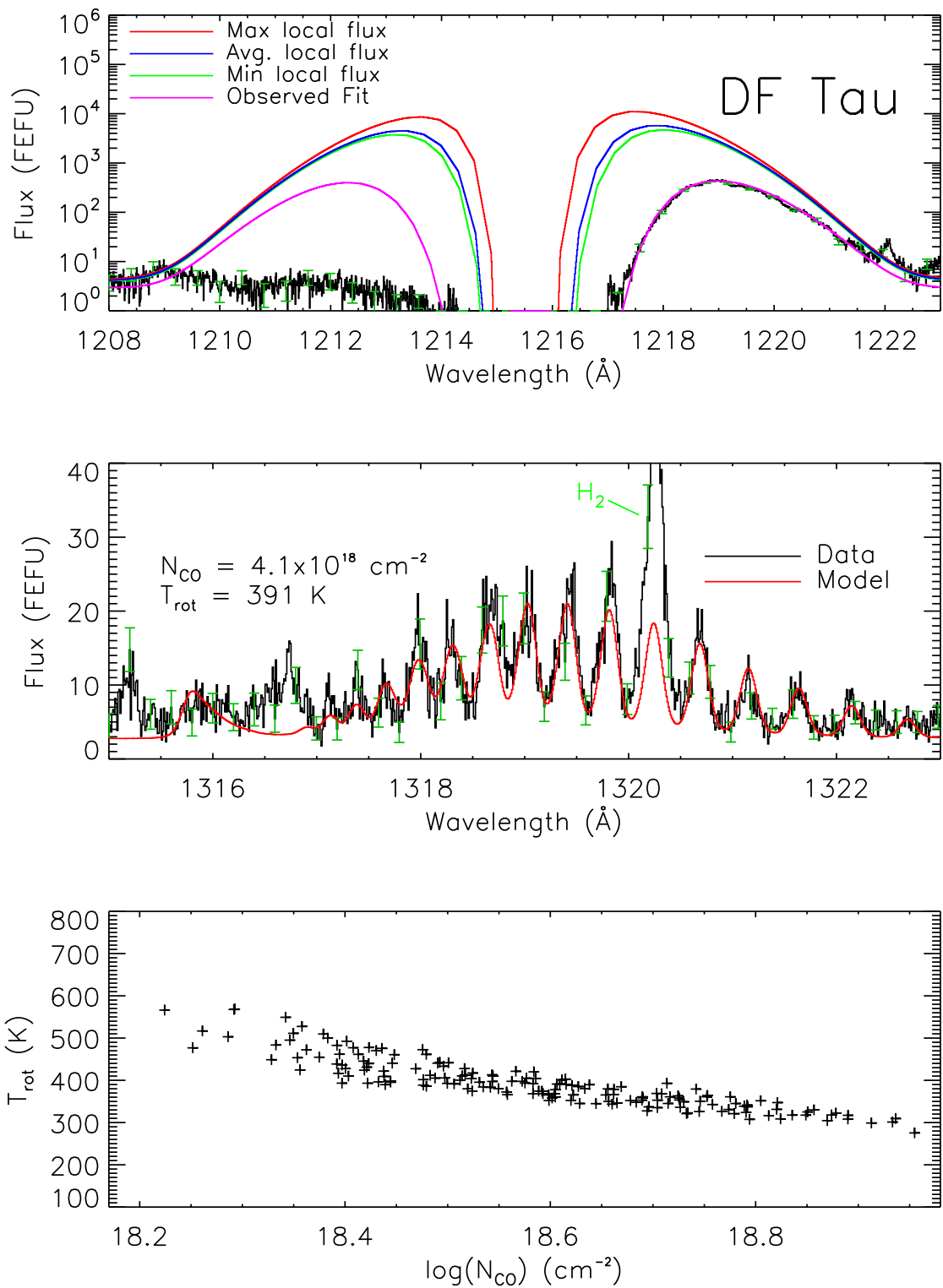}
  \caption{DF Tau model results. Top: Ly$\alpha$ data overplotted with minimum, average, and maximum model Ly$\alpha$ profiles.  Middle: CO (14 - 3) band with model from average $T_{CO}$ and $N_{CO}$.  Bottom: $T_{CO}$ vs. $N_{CO}$ for best fit models.  The red square shows the values used in the middle plot.}
   \label{fig:DFTAU_final}
  \end{figure}

The top panel of Figure \ref{fig:DFTAU_final} shows representative Ly$\alpha$ model results for DF Tau.   Several model Ly$\alpha$ profiles are plotted over the observed Ly$\alpha$ profile.  The red, blue, and green lines correspond to the maximum, average, and minimum Ly$\alpha$ emission profiles after being absorbed by the outflow H I of each respective model.  These are displayed to convey the variety of incident radiation profiles that, when used in the CO fluorescence model, produce CO emission that matches the data.  The magenta line shows a typical outflow + ISM-absorbed profile that fits the Ly$\alpha$ profile.  In the case of DF Tau the entire blue wing is absorbed, which we attribute to a non-gaussian absorber velocity profile; therefore we only fit the red wing.  The blue wing is generally fitted between 1210 \AA\ and 1214 \AA, while the red wing is fitted from 1217 \AA\ to 1220 \AA.   We did not use two Ly$\alpha$ emission components for DF Tau and RECX 11, as a second component did not significantly improve the fit.  Strong H$_2$ absorption exists in the Ly$\alpha$ wings of V4046 Sgr, DF Tau, RECX 15, and DM Tau \citep{Yang2011}; a spline interpolation of the profile was used to minimize their influence on the fits.  The H$_2$ absorption may be present in the Ly$\alpha$ profile incident on the fluoresced CO, but the absorption has a negligible effect on the resulting CO fluorescence.

We searched a grid of the parameters to find the best-fitting combinations.  A ``good" fit was determined by minimizing $\chi^2$.  Fits were excluded with $\chi^2$ greater than 95.4 $\%$ (2$\sigma$) from the peak of the $\chi^2$ probability distribution.  In Tables \ref{TBL:Lyman_parameters_1} and \ref{TBL:Lyman_parameters_2} we list parameter values of the best-fitting Ly$\alpha$ profiles.  Since our model Ly$\alpha$ profiles must not only fit the observed Ly$\alpha$ emission but also produce the appropriate CO emission through our fluorescence model, we only include models that create CO emission that fits the (14 - 3) band.  The Ly$\alpha$ parameters listed are determined by taking the average and standard deviation of each parameter for all models.  The broad range of values for the emission components reflects the variety of emission profiles that in combination with different absorbing profiles match the data.  Continuum values are determined by averaging emission-free ranges of the surrounding spectrum.  The ``outflow'' absorber velocites for DM Tau are positive, possibly due to predominantly inflowing material.  Overall, our combined values for $N_{ISM}$ and $N_{out}$ are similar to those measured previously in CTTSs \citep{Walter2003,Herczeg2006}.

\subsubsection{CO Model Fits}

The middle panel of Figure \ref{fig:DFTAU_final} shows a typical CO model fit to the (14 - 3) band, along with its value for $N_{CO}$ and $T_{CO}$.  However, this $N_{CO}$ and $T_{CO}$ combination is the best fit for only one of our model Ly$\alpha$ profiles.  The bottom panel shows the distribution of best fit $T_{CO}$ vs. $N_{CO}$ for all the best fit Ly$\alpha$ profiles, to demonstrate the range of possibilities enabled by our model.  Each point corresponds to a unique set of Ly$\alpha$ profile, $N_{CO}$, and $T_{CO}$ that in combination adequately reproduce the observed Ly$\alpha$ and CO emission.  The error bars for each fit are not included, as they are smaller than the overall distribution of points.  Ly$\alpha$ and CO must be significantly detected in order to sufficiently constrain $N_{CO}$ and $T_{CO}$; even for the targets with the highest quality Ly$\alpha$ and CO profiles (V4046 Sgr, DF Tau, and RECX 15), the uncertainty in the Ly$\alpha$ incident radiation causes a spread in the best-fitting $N_{CO}$ and $T_{CO}$.

For fitting the (14 - 3) band we choose the regions between 1315 \AA\ and 1322 \AA\ carefully so as to avoid the 1316.55 and 1320.22 \AA\ H$_2$ lines.  We include the bandhead ($J''$ $<$ 5), as  it constrains $N_{CO}$ for low $T_{CO}$.  Also, the high $J$ lines 1317 to 1320 \AA\ must be included, as they constrain $T_{CO}$ for a given value of  $N_{CO}$.  The range between 1316.5 and 1317.5 \AA\ is generally excluded; interstellar H I absorption removes any knowledge of the Ly$\alpha$ flux which results in CO emission in this region.  This corresponds to the $J'$=6-9, 9-12, and 12-15 transitions for the P, Q, and R branches, respectively.  For the (14 - 4) band in V4046 Sgr, RECX 15, and DM Tau, we exclude flux between 1353.5 \AA\ and 1354.5 for the same reason.  Due to the computation time necessary to perform a grid search with our fluorescence model, we utilized a parameter optimization program starting from an initial coarse parameter grid.

Table \ref{TBL:CO_parameters} lists the CO fit parameters for all targets.  Individual confidence intervals for each fit are smaller than the overall distribution of good fits.   For this reason, the CO parameters we list are determined from the minimum and maximum parameter values, with the average of the two chosen as the mean value.  Information for the (14 - 4) fits is also included for the three targets free of O I ] emission (V4046 Sgr, RECX 15, and DM Tau).  For  targets with both (14 - 3) and (14 - 4) bands fits, the values of $T_{CO}$, $N_{CO}$, and $\Delta v_{broad}$ agree within error bars.  The required $v_{CO}$ to fit each band differs by $<$10 km s$^{-1}$, which is below the accuracy of the COS wavelength solution.  The lack of knowledge about the core of the Ly$\alpha$ profile is chiefly responsible for discrepancies between the model flux and that of the observed spectrum.  Flux in the 1316 - 1317 \AA\ region is typically under-predicted, in part due to the contaminating H$_2$ line, but also due to a more complicated absorbing outflow velocity distribution than we can reasonably model at present.  Future work will include more complicated velocity profiles.  

In DF Tau, RECX 15, and DM Tau, the high-$J$ lines appear close to unresolved, which help to strongly constrain $\Delta v_{broad}$.  For DM Tau, HD 135344B, and DE Tau, the Ly$\alpha$ and/or CO S/N is low enough to make it difficult to constrain parameters as well as in the other targets.  In RECX 11 the CO high-$J$ lines are not resolved, although whether this is physical (due to velocity broadening of close-in CO) or merely due to the lower S/N in the band is not clear.  As a result, $\Delta v_{broad}$ has a wide range of values.  DE Tau has a wide range of $T_{CO}$ and $N_{CO}$, primarily due to its poorly-constrained Ly$\alpha$.  In GM Aur, HN Tau, DR Tau, DK Tau, and LkCa 15, the highly-absorbed Ly$\alpha$ and/or weak CO emission makes it impossible to constrain $N_{CO}$ and $T_{CO}$ better than several dex.  Upper limits on $N_{CO}$ cannot even be determined without better knowledge of their Ly$\alpha$ profiles.  This will be possible in future work by partially reconstructing the Ly$\alpha$ profiles with measured H$_2$ fluxes.  The models presented here assume isothermal CO. Relaxing this assumption would provide more flexibility in simultaneously fitting low-J and high-J emission lines. The COS observations of some TTS may have enough sensitivity to test disk models that predict a range of gas temperatures.

\section{Discussion}

\subsection{Spatial Distribution of UV-CO}

While the degeneracy of $N_{CO}$ and the filling fraction prevents us from definitively determining the geometry of the UV-CO around the star, comparing our measured UV-CO properties with those of previously measured disk gas still reveals much about the spatial distribution of the gas.  CO M-band fundamental emission ($v'$=1, $v''$=0) has been detected in several of these targets \citep{Salyk2009,Najita2003, Bast2010,Pontoppidan2008}. The gas in which the IR-CO emission originates is typically warm, thermally excited at temperatures as high as $\sim$1500 K.  Excitation diagrams of larger IR-CO samples can be adequately described by LTE models of a single-temperature slab of gas \citep{Salyk2011}. Line-width analysis suggests that this emission is produced $\leq$ 1 AU from the central star. This information, combined with our measured UV-CO properties, presents two arguments that the UV-CO and IR-CO are spatially separate.

The UV-CO temperatures suggest separate populations, as our measured $T_{CO}$ values are several times lower than typical IR-CO temperatures. The ro-vibrational levels of the CO ground electronic state are primarily populated based on the Einstein A coefficients of each $v$ and $J$ level, relative to the temperature-dependant collision rates and gas density. Models typically assume the IR-CO lies in a "skin" of the gas disk where densities are high enough for the rotational temperature to be equal to the kinetic temperature \citep{Brittain2007}.  If this is also true for the UV-CO, then it must lie in a different region of the disk where the gas kinetic temperature is lower.  However, as temperature and density vary based on location in the disk it may be possible to find gas populations with thermal low-$J$ levels, while the high-$J$ levels are subthermal. In targets with well-defined high-$J$ lines (V4046 Sgr, DF Tau, RECX 15, and DM Tau), these lines strongly influence the $T_{CO}$ and $N_{CO}$ fits. If densities are less than the critical densities for these high-$J$ states ($n_{crit}$ for $J$=18 at 400 K is $\sim10^6$ cm$^{-3}$), then the derived temperature could be less than that determined from the thermalized lower states. In this case the CO density would have to be lower than the thermalized IR-CO, and therefore be in a different region of the disk.

Our measured $\Delta v_{broad}$ values enable us to spatially constrain the UV-CO for the targets with well-defined high-$J$ lines.  Due to the blending of the UV-CO bands we cannot calculate the line width at zero intensity of individual $J$ transitions, from which the inner CO radius is typically calculated \citep{Brittain2009}.  However, assuming that the FWHM of the high-$J$ lines is primarily due to orbital motion of the gas, we can calculate an average orbital radius $R_{CO}$: 

\begin{equation}
R_{CO} = GM_*\left(\frac{2sin(i)}{\Delta v_{broad}}\right)^2
\label{radius}
\end{equation}

\noindent where $G$ is the gravitational constant, $M_*$ the stellar mass, and $i$ the disk inclination, previously measured in the literature.  For the values listed in Appendix \ref{targets}, we calculate $R_{CO}$ for V4046 Sgr, DF Tau, RECX 15, and DM Tau: 2.76 AU, 2.16 AU, $>$3.04 AU, and $>$1.56 AU respectively.  These radii are outside typical IR-CO radii ($R_{CO}$ $<$ 1 AU) calculated with Equation \ref{radius} and FWHM values from a survey of CTTSs from \citet{Salyk2011b}.  DF Tau is observed in their survey, with $R_{CO}$ = 0.15 AU based on their CO FWHM (79 km s$^{-1}$), inclination (85$^{\circ}$,  \citet{JKrull2001}), and M$_*$ (0.27 M$_{\odot}$, \citet{Gullbring1998}). The $\Delta v_{broad}$ values of RECX 15 and DM Tau are at the COS resolution, so we can only determine lower limits for $R_{CO}$ in these disks.  If thermal broadening, turbulent velocity, or disk winds contribute to the line widths, then R$_{CO}$ could be even larger.  Regardless, the temperature and $\Delta v_{broad}$ disagreements lead us to the conclusion that the UV-CO and IR-CO are indeed separate populations.

In addition to the spatial differences between the two CO populations, there is no consistent agreement between the detection of IR-CO and UV-CO emission in our sample.  DR Tau exhibits strong IR-CO emission \citep{Salyk2008}, but no UV-CO emission is observed.  Conversely, DM Tau shows strong UV-CO emission, but no IR-CO emission.  DF Tau, HD 135344 B, and DE Tau show both.  Unfortunately, the stronger detection targets V4046 Sgr, RECX 15, and RECX 11 have yet to be observed in the M-band.  Different morphology (UV-CO only, IR-CO only, both) is not a strong argument for separate gas populations. In principle, different morphologies could occur in a single gas population if the Ly$\alpha$ pumping and IR excitation mechanisms are not physically related.  Further study of a larger survey will be necessary to draw any solid conclusions on this matter, but the temperature and radius differences found in our analysis argue against a single gas population.

The UV-CO also appears to differ from H$_2$ previously studied in CTTSs.  H$_2$ is best observed in the UV due to its lack of a permanent dipole moment.  In the TW Hya disk, \citet{Herczeg2004} detected H$_2$ emission, measuring $T_{H_2}$ = 2500$^{+700}_{-500}$ K and log$N_{H_2}$ = 18.5$^{+1.2}_{-0.8}$ cm$^{-2}$.  In addition, France et al. (2011, in press) observed H$_2$ in absorption in the spectrum of AA Tau, measuring $T_{H_2}$ = 2500$^{+800}_{-700}$ K and log$N_{H_2}$ = 17.9$^{+0.6}_{-0.3}$ cm$^{-2}$.  As with the IR-CO, the differences in temperature suggest that the two populations reside in different regions of the disk.  Similar to the conclusion of France et al. (2011, in press) about the AA Tau CO and H$_2$, a subthermal CO population could explain the disparity. If the CO and H$_2$ are indeed cospatial, then the similarity of $N_{CO}$ and $N_{H_2}$ derived from UV spectra would seem to violate the CO/H$_2$ ratio of 10$^{-4}$ \citep{Lacy1994}.  It is possible that during the process of star and planet formation, the CO is enhanced and/or H$_2$ removed from the gas phase in a way that increases the CO/H$_2$ ratio.  As suggested by \citet{France2011b}, collisions of large grains or planetesimals may re-release gas-phase CO into the disk.  In addition, H$_2$ could be selectively dissociated by radiation in the inner disk.  However, in order to truly compare both molecules the UV H$_2$ and CO emission must be studied with a self-consistent approach.

\begin{figure}
  \centering
  \includegraphics[width=15cm]{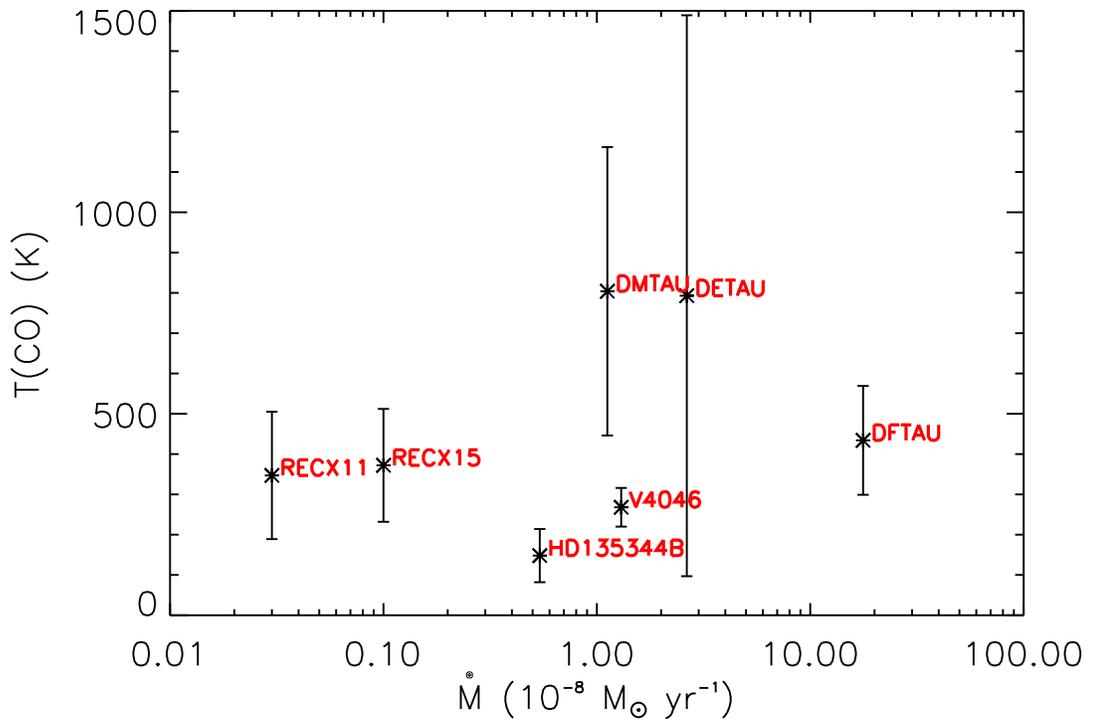}
  \caption{Model-derived $T_{CO}$ vs. literature accretion rates. }
   \label{fig:N_T_Mdot}
  \end{figure}

Comparing $N_{CO}$ and $T_{CO}$ with other disk properties published in the literature does not yield any significant correlations.  For example, an increase in rotational temperature vs.\ accretion rate might be expected; higher accretion rates create greater FUV continuum and Ly$\alpha$ emission, which in turn heats the gas to higher temperatures via photo-electric heating.  The FUV photons could also populate higher $J$ states in a way not possible by collisional processes alone.  However, as seen in Figure \ref{fig:N_T_Mdot}, there appears to be no correlation between temperature and accretion rate.  Over more than two orders of magnitude of accretion rate, targets with strong CO and Ly$\alpha$ emission well constrain the temperature between 200 K and 500 K with no apparent trend.  Only in DM Tau and DE Tau is the temperature less constrained, but this is primarily due to their highly absorbed Ly$\alpha$ profiles.

\subsection{CO Band Flux and Disk Evolution}
\label{fluxes}

\begin{figure}
  \centering
  \includegraphics[width=15cm]{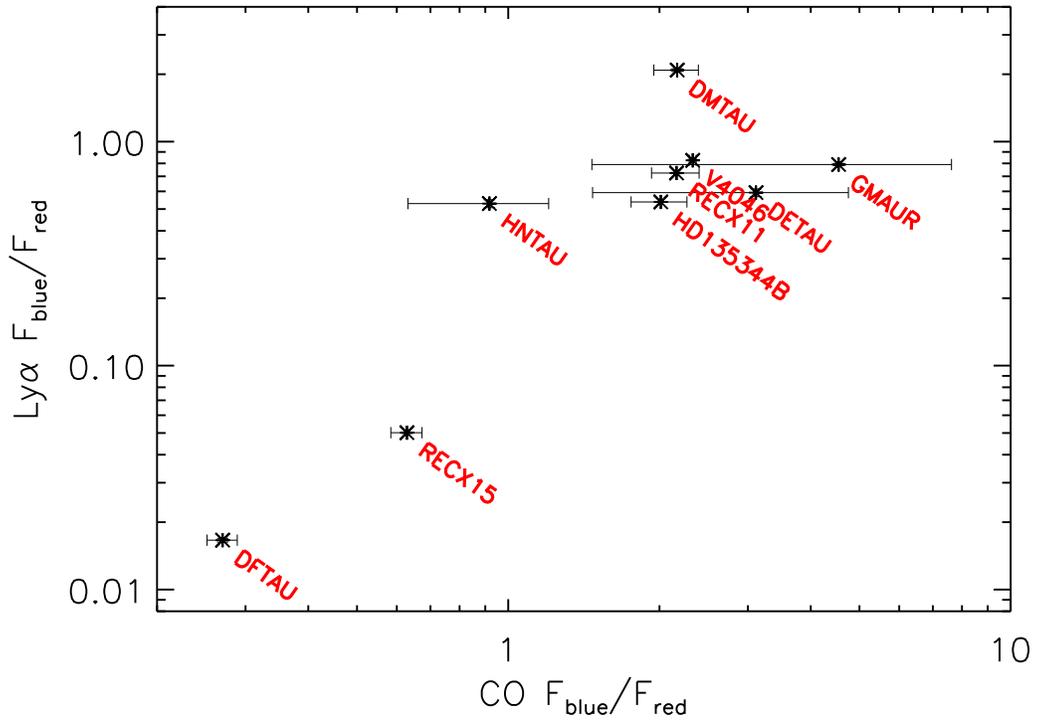}
  \caption{Ratios of blue to red flux for Ly$\alpha$ and CO.  The blue and red Ly$\alpha$ fluxes were summed from 1210 - 1214.7 \AA\ and 1216.7 - 1222 \AA, respectively.}
   \label{fig:Flux_ratios}
  \end{figure}

The total CO emission from our sample can be evaluated by integrating the (14 - 3) band flux.  Referring to Figures \ref{fig:Lyman_1315} and \ref{fig:Lyman_1315_non}, the (14 - 3) bands are summed between the dashed grey lines at 1315.2 \AA\ and 1320 \AA.  The continuum level $C_{(14 - 3)}$ is determined from the surrounding line-free spectrum, and denoted by the horizontal solid black/blue/red line.  The blue and red lines identify the low and high $J$ state emission regions, which are separately integrated.  The dashed orange lines above and below mark the 1-$\sigma$ error of the continuum level.  The integrated flux values are listed in Table \ref{TBL:1315_Sum}.  For DR Tau and LkCa 15 we list zero flux, which actually corresponds to an overall negative flux.  Shown in Figure \ref{fig:Flux_ratios} is ratio of Ly$\alpha$ blue to red flux vs. the ratio of CO blue (low-$J$) to red (high-$J$) flux.  This demonstrates our initial correlation between the relative strength of the Ly$\alpha$ photo-exciting radiation and the resulting CO emission.

\begin{figure}
  \centering
  \includegraphics[width=15cm]{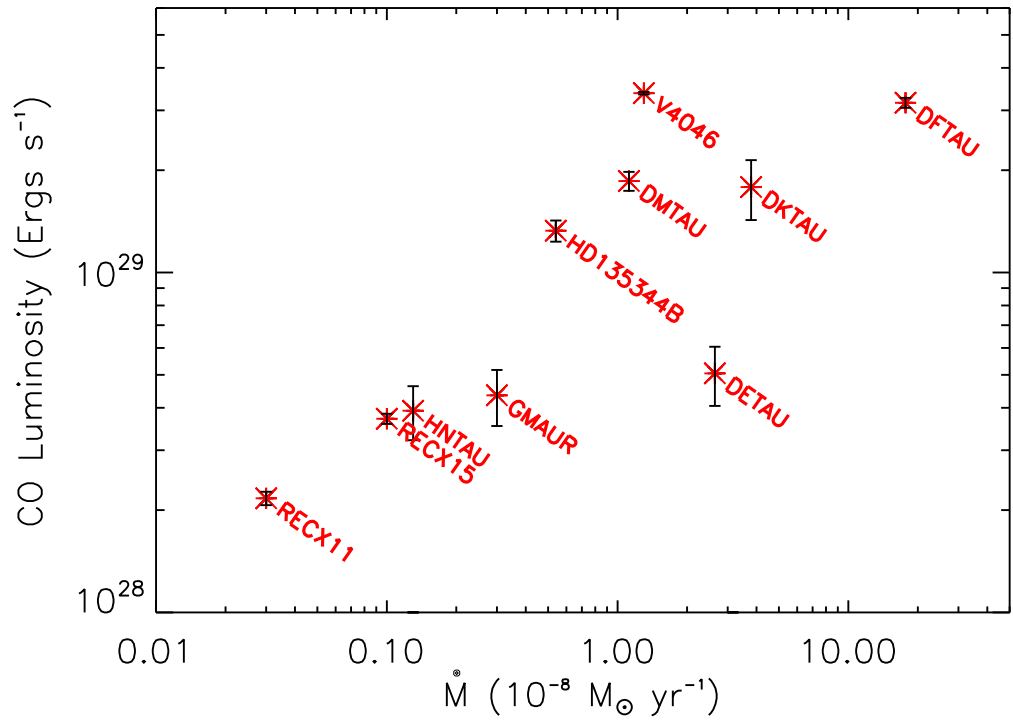}
  \caption{CO (14 - $v''$) band luminosity vs. literature accretion rates for positive detection targets.  The CO luminosities were calculated by applying branching ratios to the (14 - 3) luminosities.}
   \label{fig:CO_H2_mdot}
  \end{figure}

To test correlations with disk evolutionary properties we use the total CO (14 - $v''$) luminosity, calculated from the (14 - 3) band fluxes:

\begin{equation}
L(CO)_{14-3} = \frac{F(14 - 3)}{B(14 - 3)}4\pi d^2
\end{equation}

\noindent where F(14 - 3) is the integrated band flux, B(14 - 3) is the branching ratio listed in Table \ref{TBL:Band_Ratios}, and $d$ is the distance to the target.  In Figure \ref{fig:CO_H2_mdot}, $L(CO)_{14-3}$ shows a strong correlation with mass accretion rate.  This is not surprising, as the amount of photo-exciting Ly$\alpha$ is expected to increase with accretion.  Combined with the weak opacity of the CO (14 - 0) band, we are simply seeing the re-emission of most of the Ly$\alpha$ photons capable of being absorbed.  

\begin{figure}
  \centering
  \includegraphics[width=15cm]{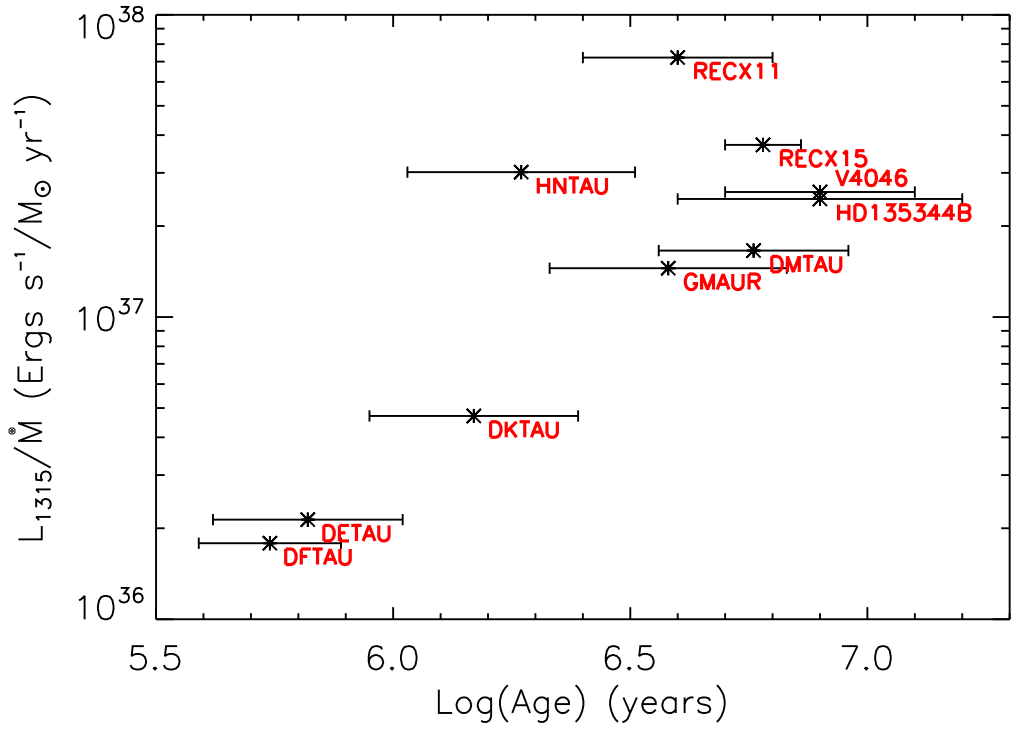}
  \caption{CO (14 - $v''$) band luminosities, divided by literature accretion rates, demonstrate an increasing correlation with age.}
   \label{fig:CO_age}
  \end{figure}

As a result, $\dot M_{acc}$ cannot be used as a proper evolutionary diagnostic, despite the fact that accretion does decrease as a CTTS ages \citep{Hartmann1998,Fang2009}.  We attempt to correct for this by dividing the CO luminosity by $\dot M_{acc}$ to compare with age. Figure \ref{fig:CO_age} shows this relation, with a trend towards more CO emission in more evolved systems.  One interpretation is that as inner disks clear, the Ly$\alpha$ can more readily illuminate the CO.  V4046 Sgr, HD135344B, DM Tau, and GM Aur are all confirmed transitional objects with cleared dust holes in the center of their disks \citep{Jensen1997,Rodriguez2010,Brown2009, Calvet2005}, which would allow the Ly$\alpha$ photons to reach more CO.  CO-IR emission from HD 135344B and GM Aur support this argument, as the CO-IR appears inside the inner dust radii \citep{Pontoppidan2008}. However, no UV-CO is detected in the transitional disk LkCa 15 A second interpretation would be the increase of gas-phase CO through grain and planetesimal collisions \citep{France2011b}. CO produced in the dust disk could be transported to the inner disk surface as the system evolves.

\subsection{C IV Photoexcitation}
\label{civfluxes}

\begin{figure}
  \centering
  \includegraphics[width=14cm]{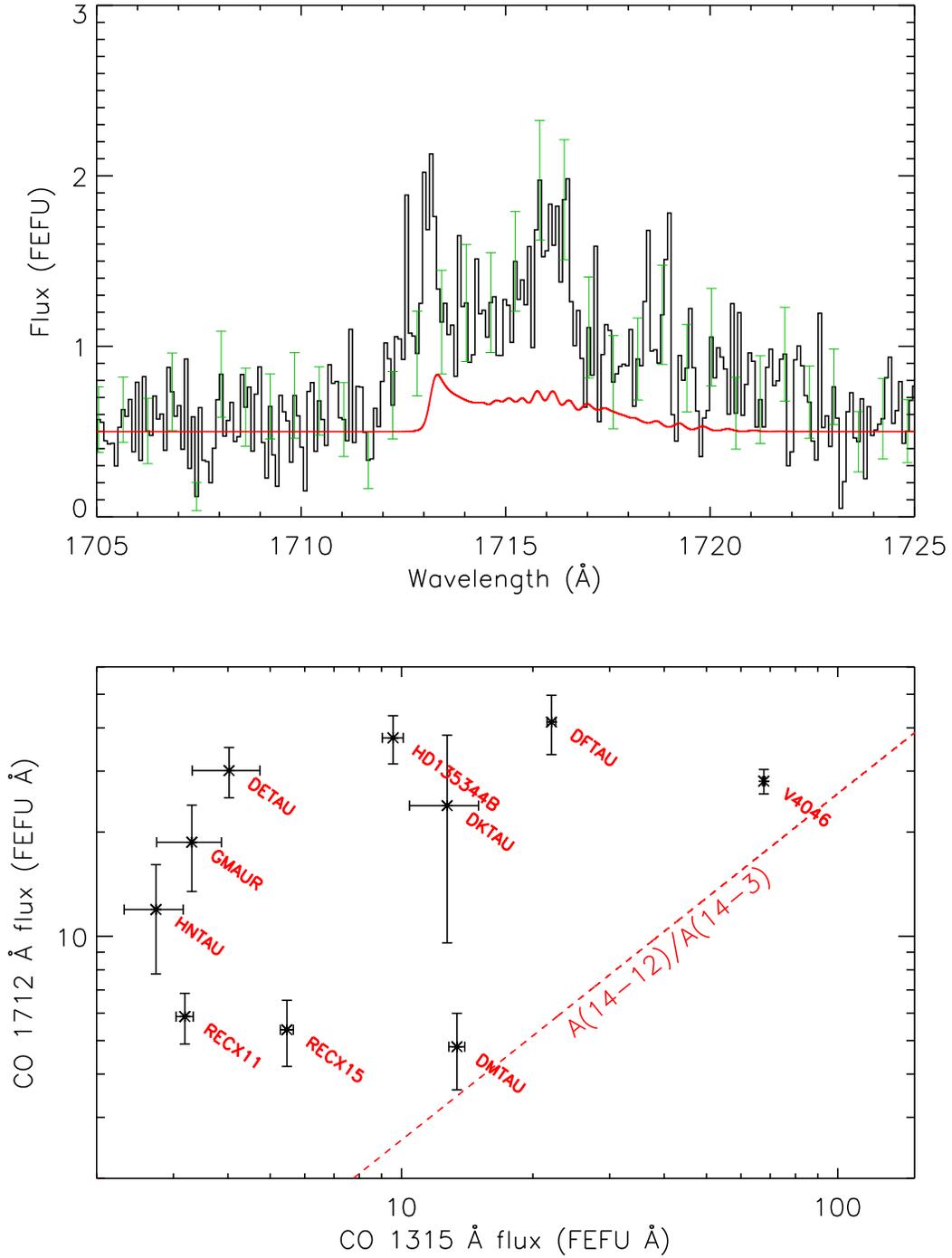}
  \caption{Top: RECX 11 (14 - 12) model emission (red) over the observed emisison.  The data have been smoothed to one spectral resolution element for display purposes.  Summed fluxes for the 1712 \AA\ emission vs. the 1316 \AA\ band, for each target.  The dashed red line shows the expected (14 - 12) flux expected from branching ratios for a given (14 - 3) flux.}
   \label{fig:CIV_pumped}
  \end{figure}

Finally, we measure flux in the 1712 - 1720 \AA\ region, assuming that emission in this region is due to the (14 - 12) band.  Of particular note is how large these fluxes are compared to what would be expected from simply applying branching ratios to the (14 - 3) fluxes.  Figure \ref{fig:CIV_pumped} shows this graphically; for one of the best-fit (14 - 3) models of RECX 11, the corresponding (14 - 12) emission is plotted over its observed 1712 \AA\ emission.  This trend is seen significantly in all targets and also shown in the bottom of Figure \ref{fig:CIV_pumped} with the dashed line representing the expected (14 - 12) integrated flux for a given (14 - 3) integrated flux.  If this emission is due to CO, some process other than Ly$\alpha$ photo-excitation must be responsible.  It is likely the result of C IV photo-excitation, which is capable of exciting the $v'$=0 level of CO \citep{France2011b}.  The resulting (0 - 3) transition creates a band of emission at 1712 - 1720 \AA.  Cursory analysis shows that there is sufficient flux in the observed C IV lines to produce these levels of emission.

\section{Conclusions}

In this paper, we have characterized UV photo-excited CO emission in a sample of T Tauri star spectra from the COS instrument on $HST$, reaching the following conclusions:

1.  Electronic transitions of CO are observed in approximately half of the T Tauri stars surveyed, primarily due to photo-excitation by Ly$\alpha$ emission coincident with the $A^1\Pi$ - $X^1\Sigma^+$ (14 - 0) absorption band.  The resulting transitions from the $v'$ = 14 state appear as emission bands between 1280 \AA\ and 1712 \AA.  The (14 - 3) and (14 - 4) bands are most conducive to spectral modeling.

2.  The observed Ly$\alpha$ emission is not sufficient to reproduce the observed CO band emission; thus, we create a model to simulate the local Ly$\alpha$ radiation field incident on the CO molecules.  With two Gaussian emission components created at the star, followed by outflow and ISM absorbers, we are able to recreate the observed Ly$\alpha$ profiles and CO band emission.

3.  For seven of the twelve targets, we measure CO temperatures between $\sim$100 and 1500 K, and CO column densities of $\sim10^{18}$ - 10$^{19}$ cm$^{-2}$.  Both strong Ly$\alpha$ and CO emission are required in order to constrain $N_{CO}$ to within 0.5 dex; for these 5 targets temperatures are below 600 K.  The high-$J$ lines in several targets are narrow enough to constrain the CO to outside 1 AU.

4.  There are no strong correlations between our calculated $N_{CO}$ and $T_{CO}$ values and other previously determined parameters.  Our measured CO temperatures are lower than previously measured IR-CO and UV-H$_2$ in CTTSs.  The temperature differences, along with our calculated emission line widths, imply that the UV-CO and IR-CO are different populations.  

5.  If the UV-CO and UV-H$_2$ are co-spatial, our $N_{CO}$ values are 3-4 orders of magnitude greater than those expected from previously measured H$_2$ abundances and dense cloud CO/H$_2$ ratios.  Further analysis of the H$_2$ lines, as well as more detailed geometric modeling, is essential to understand the relationship between the UV-CO, IR-CO and H$_2$

6.  CO emission increases with $\dot M_{acc}$ because of the corresponding increase in the Ly$\alpha$ radiation field.  Upon correcting for $\dot M_{acc}$, there is still a correlation between CO emission and age.  We conclude that as CTTSs evolve and disk dust and H I clears, the Ly$\alpha$ photons are able to reach more CO.

\acknowledgments
The authors thank Jeff Linsky and the anonymous referee for their helpful suggestions that improved the quality of the paper. This work was supported by NASA grants NNX08AC14G, NNX08AX05H, and NAS5-98043 to the University of Colorado as well as STScI grants to program GO-11616 and GO-11828.

\appendix

\section{Targets}
\label{targets}
\subsection{$\eta$ Chamaeleontis Targets}

The $\eta$ Chamaeleontis cluster is located at a distance of 97 pc.  Age estimates of its constituent stars range from 4 - 10 Myr \citep{Lawson2001}.  It is unusual compared to other star forming regions;  \citet{Sicilia2009} measured a disk fraction of $\sim$50$\%$, much higher than expected for an older region.  There is also a higher percentage of disk systems in transition objects than found in other clusters \citep{Megeath2005}.  

\subsubsection{RECX 11}
RECX 11 is a weakly accreting CTTS.   Its low accretion rate ($\dot M = 4$$\times$$10^{-11}M_{\odot} yr^{-1}$) would imply that photoevaporation should have cleared the inner disk, but this is not seen \citep{Gautier2008}.  UV-H$_2$ emission implies the presence of gas \citep{France2011a}.  \citet{Ingleby2011} also find low accretion values ($\dot M = 3$$\times$$10^{-10}M_{\odot} yr^{-1}$) using the same data set presented here and argue that negligible planet formation has occured during the disk evolution.  From H$_2$ line profiles, they measure rotational velocities fast enough to place the molecular gas as close as 3 R$_*$.

\subsubsection{RECX 15}

RECX 15 is an active accretor; the H$\alpha$ EW is 90.0 \AA, and its profile is indicative of a blue-shifted wind component \citep{Lawson2004}. The SED implies a continuous disk from close to the stellar surface out to tens of AU or more, where the disk flares strongly \citep{Gautier2008,Woitke2011}. However, the H$_2$ (1 - 0) S(1) emission line width measured by \citet{RHowat2007} places the UV-H$_2$ a few AU from the star.  RECX 15 has a stellar mass of 0.2 M$_{\odot}$ and inclination angle of 40$^{\circ}$ \citep{Woitke2011}.

\subsection{Taurus-Auriga Targets}
\label{taurus}

The Taurus-Auriga complex is one of the nearest star forming regions at a distance of $\sim$140 pc.   Its paucity of O/B stars, combined with a low interstellar radiation field, enable the persistence of many disks \citep{Luhman2010}.  Disk fractions in the Taurus region have been measured from $\sim$ 45\% for low-mass objects (M=0.01-0.3 M$_{\odot}$) to $\sim$75\% for solar-mass stars.   Some of these objects have been studied in exceptional detail, while others have only been measured in large surveys. 

\subsubsection{DM Tau}

DM Tau is considered to be a transitional disk due to its flux deficit shortward of 8 $\mu$m \citep{Calvet2005}. Resolved images and spectra of dust continuum emission show an inner hole of 19 AU with little dust inside this region (0.0007 lunar masses), although with a detectable accretion rate ($\sim$1.3$\times$10$^{-8} M_{\odot} yr^{-1}$) there must be gas present \citep{Calvet2005,Andrews2011}.  The disk exhibits a rich chemistry; organic molecules CO, CN, CS, HCN, HNC, HCO$^+$, C$_2$H, and N$_2$H$^+$ have been measured in emission \citep{Dutrey1997}.  The abundances of these molecules show that stellar UV photons have strong influence over their creation and destruction.  DM Tau has a stellar mass of 0.43 M$_{\odot}$ and inclination angle of 35$^{\circ}$ \citep{Hartmann1998,Andrews2011}.

\subsubsection{HN Tau}

HN Tau is a binary system (0.7 and 0.4 M$_{\odot}$) with strong jets that expand with distance from the primary star \citep{Hartigan2004}.  In this work we study only the spectrum of HN Tau A, due to its separation from HN Tau B (3.1").   M-band observations from NIRSPEC on Keck reveal weak CO emission, but strong absorption (J.M. Brown, private communication).  While its exact inclination is not known, \citet{France2011b} measure CO electronic transitions detected in absorption, which could be indicative of disk gas viewed at high inclination.  

\subsubsection{DR Tau}

DR Tau is a K4 star with an accretion rate that appears to vary by over two orders of magnitude \citep{Gullbring2000,JKrull2002}.  DR Tau shows strong molecular emission at IR wavelengths.  Rotational transition emission lines of H$_2$O are seen, in high $J$s (20-50), as well as CO and OH at temperatures around 1000 K \citep{Salyk2008}.  By modeling the gas as an isothermal LTE ring, the emission radii are 1.0, 0.8, and 0.7 AU for H$_2$O, CO, and OH, respectively.  Time-lapsed CO measurements show variability between overtone emission and absorption \citep{Lorenzetti2009}.

\subsubsection{DK Tau}

DK Tau is a binary system consisting of a K9 and M0 with $\sim$350 AU separation. From mm imaging both appear to have individual circumstellar disks \citep{Jensen2003}, however our observation was targeted on DK Tau A.  Based on the UV Unlike the majority of binary systems in which both stars contain disks, the A and B disks may be misaligned \citep{Kraus2009}.  Some molecular emission has been detected in the DK Tau disks, particularly  sub-mm $^{13}$CO emission \citep{Greaves2005}.  

\subsubsection{DF Tau}

A binary pair (M1 + M3.5) with 13 AU separation, DF Tau may be as young as 0.1 Myr \citep{Ghez1997}.  Our spectrum samples the primary, as the secondary is very dim at UV wavelengths. Strong accretion has been measured from a UV continuum \citep{Herczeg2006,France2011a}, as have UV H$_2$ emission lines.  H$_2$ appears both in Ly$\alpha$/C IV pumped emission, and in absorption in an outflow \citep{Yang2011}.   DF Tau displays a rich spectrum of IR-CO emission lines; the CO (1 - 0) R(3) shows a FWHM of $\sim$ 74 km s$^{-1}$, implying a close orbit to the star \citep{Najita2003}.  \citet{Greaves2005} also detect sub-mm emission from $^{13}$CO.  DF Tau has a mass of 0.27 M$_{\odot}$ and inclination of 85$^{\circ}$ \citep{Gullbring1998,JKrull2001}

\subsubsection{LkCa15}

LkCa15 is a pre-transitional CTTS with a ringed gap in its disk, likely caused by a (proto)planet \citep{Kraus2011}.  From mid-IR imaging the inner rim of the optically thick dust disk is calculated to be 46 AU, with some remnant dust inside \citep{Espaillat2007,Pietu2006}.   \citet{Najita2003} measured fundamental CO emission as close as 0.1 AU.

\subsubsection{GM Aur}

GM Aur is a massive transitional disk; the $Spitzer$ Infrared Spectrograph (IRS) SED is best explained by 0.02 lunar masses of dust within a 5 AU inner disk, and an outer disk with a 24 AU inner radius \citep{Calvet2005}.  High resolution sub-mm observations support this picture, finding a 28 AU hole in the optically thick dust disk \citep{Andrews2011}.  \citet{Salyk2007} measured near-IR CO emission within 0.2$^{+0.4}_{-0.2}$ AU, at high rotational temperatures, coincident with the measured dust inner rim.  Ne II was measured in the surrounding disk at the systemic velocity of 14.8 km s$^{-1}$ \citep{Najita2009}.  Along with a profile broadened by rotational velocity, this would suggest the gas lies in the disk. 

\subsection{Isolated Targets}
\label{isolated}

Two of the targets do not reside in star-forming regions.  V4046 Sagittarius is a $\sim$9 Myr old CTTS, a closely orbiting ($\sim$ 9 R$_{\odot}$ separation) binary with combined mass of 1.78 M$_{\odot}$ \citep{Stempels2004} at a distance of 83 pc \citep{Rodriguez2010}.   At an inclination of 36$^{\circ}$, the disk appears to be coplanar with the central binary stars \citep{Monin2007}.  \citet{Kastner2008} measured molecular mm emission lines from CO, HCN, CN, and HCO$^+$; the double peaked CO profiles confirm outer disk Keplerian rotation.  \citet{Rodriguez2010} imaged the system in $^{12}$CO(2-1) and $^{13}$CO(2-1) millimeter emission, measuring the molecular disk extent to $\sim$ 370 AU and mass of $\sim$ 110 Earth masses.  HD 135344B is an F star in a visual binary, and has a disk with a gap from .45 to 45 AU \citep{Brown2007}.  It is more massive than the other stars in our sample.  CO has been detected in emission inside the hole out to $\sim$15 AU \citep{Brown2009}.  The distribution of gas column densities agree with models of a Jovian planet orbiting between 10 and 20 AU \citep{Pontoppidan2008}.


\begin{sidewaystable}
 \caption{ \textit{HST}-COS observing log}
\begin{center}
\begin{tabular}{cccccc}\hline \\
Object & R.A. & Dec. & Date &  T$_{exp}$ (s) & Program\\ 
Object & (J2000) &  (J2000) &  &  (G130M,G160M) & \\ \hline
RECX 11 &  08$^h$ 47$^m$ 01.28$^s$ & -78$^{\circ}$ 59' 34.1" & 2009 December 12 &  3645, 4514 & 11616\\
DF Tau & 04$^h$ 27$^m$ 02.81$^s$ & +25$^{\circ}$ 42' 22.1" & 2010 January 11 &  4828, 5549 & 11533\\
DK Tau & 04$^h$ 30$^m$ 44.25$^s$ & +26$^{\circ}$ 01' 24.5" & 2010 February 4 &  1585, 1971 & 11616\\
LkCa15 & 04$^h$ 39$^m$ 17.79$^s$ & +22$^{\circ}$ 21' 03.1" & 2010 February 5 &  1712, 1725 & 11616\\
RECX 15 & 08$^h$ 43$^m$ 18.43$^s$ & -79$^{\circ}$ 05' 17.7" & 2010 February 5 & 3891, 4502 & 11616\\
DR Tau & 04$^h$ 47$^m$ 06.22$^s$ & +16$^{\circ}$ 58' 42.6" & 2010 February 10 &  1704, 1746 & 11616\\
HN Tau & 04$^h$ 33$^m$ 39.37$^s$ & +17$^{\circ}$ 51' 52.1" & 2010 February 10 &  5725, 4529 & 11616\\
HD135344B & 15$^h$ 15$^m$ 48.42$^s$ & -37$^{\circ}$ 09' 16.3" & 2010 March 14 &  2666, 4836 & 11828\\
V4046 Sgr & 18$^h$ 14$^m$ 10.49$^s$ & -32$^{\circ}$ 47' 34.2" & 2010 April 27 &  4504, 5581 & 11533\\
GM Aur & 04$^h$ 55$^m$ 10.98$^s$ & +30$^{\circ}$ 21' 59.1" & 2010 August 19 &  2128, 1861 & 11616\\
DE Tau & 04$^h$ 21$^m$ 55.69$^s$ & +27$^{\circ}$ 55' 06.1" & 2010 August 20 &  2068, 1851 & 11616\\
DM Tau & 04$^h$ 33$^m$ 48.74$^s$ & +18	$^{\circ}$ 10' 09.7" & 2010 August 22 &  3459, 3770 & 11616\\
 \hline
\end{tabular}
\end{center}
\label{TBL:Observing_Log}
\end{sidewaystable}

\begin{sidewaystable}
{\bf \caption{\label{Target_Parameters}Target Parameters}}\begin{center}
\begin{tabular}{cccccccc}\hline\\
Object & Sp. Type & Group & d &  A$_V$ & $\dot M $ &  log$_{10}$(Age) &References\\ 
 & & & (pc) & & (10$^{-8}$M$_{\odot}$ yr$^{-1}$) &  (yrs)\\ \hline
V4046 Sgr & K5, K7 & Isolated & 83 &  0& 1.3 &    7.08 $\pm$ 0.20 & 6,8,10\\
DF Tau & M1, M3.5 &Taurus-Aur & 140 &   0.5 & 17.69 &  5.74 $\pm$ 0.15  & 1, 11\\
RECX 15 & M3.4 & $\eta$ Cha  & 97 &   0  & 0.1 &   6.78 $\pm$ 0.08  & 5,12 \\
DM Tau & M1 & Taurus-Aur & 140 &  0.5& 1.12 &   6.76 $\pm$ 0.20  & 3,7,11\\
RECX 11 &  K6.5 & $\eta$ Cha & 97 &  0 & 0.03 &    6.60 $\pm$ 0.20  & 5,13\\
HD135344B & F4 & Isolated &140 &  0.3 & 0.54 &    6.90 $\pm$ 0.30  & 4,14\\
DE Tau & M0 & Taurus-Aur & 140 &  0.62&  2.64 &   5.82 $\pm$ 0.20  & 1,11\\
GM Aur & K7 &Taurus-Aur & 140 &  0.96 & 0.3 &   6.58 $\pm$ 0.25  & 1,11\\
HN Tau & K5, M4 & Taurus-Aur & 140 &  0.65 & 0.13   &6.27 $\pm$ 0.24  & 1,11\\
DR Tau & K4 &Taurus-Aur & 140 & 1.2 & 3.16 &   5.29 $\pm$ 0.34  & 2,9,11\\
DK Tau & K9, M0 & Taurus-Aur & 140 &  1.42 & 3.79   &6.17 $\pm$ 0.22  & 1,11\\
LkCa15 &  K5 & Taurus-Aur & 140 &  .62 & 0.13  &  6.35 $\pm$ 0.26  & 3,7,11 \\
 \hline
\end{tabular}
\\ \rule{0mm}{5mm}
(1) \citet{Gullbring1998}; (2) \citet{Gullbring2000}; (3) \citet{Kenyon1995}; (4) \citet{Garcia2006}; (5) \citet{Lawson2004}; (6) \citet{Stempels2004}; (7) \citet{Hartmann1998};  (8)  \citet{France2011a}; (9) \citet{White2001}; (10) \citet{Quast2000}; (11) \citet{Kraus2009}; (12) \citet{RHowat2007}; (13) \citet{Lawson2001}; (14) \citet{VanBoekel2005}
\end{center}
\end{sidewaystable}

\begin{table}
\begin{center}
{\bf \caption{\label{TBL:Band_Ratios}CO Fourth Positive bands identified in sample CTTSs.}}\begin{center}
\begin{tabular}{cccccc}\hline\\
Band ID &  $\lambda_{obs}$ (\AA)\tablenotemark{a} & Pumping Source & $\lambda_{pump}$ (\AA) & A($v'$ - $v''$) (s$^{-1}$)\tablenotemark{b} & B($v'$ - $v''$)\tablenotemark{c} \\ \hline
  & & & \\[-.1in]
(14 - 2) &      1280.5 &       H I Ly$\alpha$ & 1214 - 1222 & 4.0$\times$10$^8$ & 0.04 \\[.05in]
(14 - 3)&      1315.7 &       H I Ly$\alpha$ & 1214 - 1222 & 1.5$\times$10$^9$ & 0.17\\[.05in]
(14 - 4) &      1352.4 &       H I Ly$\alpha$ & 1214 - 1222 & 2.7$\times$10$^9$ & 0.30\\[.05in]
(14 - 5) &       1390.7 &        H I Ly$\alpha$ & 1214 - 1222 & 1.6$\times$10$^9$ &0.17\\[.05in]
(14 - 7)&       1472.6 &        H I Ly$\alpha$ & 1214 - 1222 & 9.5$\times$10$^8$ &0.10\\[.05in]
(14 - 8) &       1516.3  &        H I Ly$\alpha$ & 1214 - 1222 & 5.1$\times$10$^8$ & 0.06\\[.05in]
(14 - 10) &       1610.1 &        H I Ly$\alpha$ & 1214 - 1222 & 5.6$\times$10$^8$ &0.06\\[.05in]
(14 - 12) &      1713.2 &        H I Ly$\alpha$ & 1214 - 1222 & 3.9$\times$10$^8$ & 0.04\\[.05in]
(0 - 1) &       1597.3 &       C IV & 1544 - 1551 & 4.3$\times$10$^9$ & 0.33\\[.05in]
(0 - 2) &       1653.2 &       C IV & 1544 - 1551 & 3.6$\times$10$^9$ & 0.27\\[.05in]
(0 - 3) &       1712.4 &       C IV & 1544 - 1551 & 1.8$\times$10$^9$ & 0.14\\[.05in]
 \hline
\end{tabular}
\end{center}
\tablenotetext{a}{CO wavelengths are taken from Kurucz (1993).}
\tablenotetext{b}{Summed A values for each ($v'$ - $v''$) band.}
\tablenotetext{c}{Branching ratios for each vibrational band, relative to all ($v'$ - $v''$) bands.}
\end{center}
\end{table}

\begin{sidewaystable}
{\bf \caption{\label{TBL:Lyman_parameters_1}Best-fit Ly$\alpha$ parameters}}
\begin{center}
\begin{tabular}{ccccccc}\hline\\
 & I$_1$(0)& $\sigma_1$  & $v_2$   &  I$_2$(0)  & $\sigma_2$ &  $C_{Ly\alpha}$ \\ 
Object  & (FEFU) & (km s$^{-1}$) & (km s$^{-1}$) & (FEFU) &  (km s$^{-1}$)  & (FEFU) \\ \hline
V4046 Sgr  &      10122 $\pm$       3531 &        263 $\pm$         40 &        -72 $\pm$         35 &       5232 $\pm$       2840 &        368 $\pm$         31 &          7.0 $\pm$          3.0 \\ 
DF Tau  &      52084 $\pm$        943 &        374 $\pm$          0 &  - & - & - & 4.5 \\
RECX 15  &         16 $^{+  29}_{-16}$ &        265 $^{+292}_{-265}$ &        208 $\pm$         28 &        906 $\pm$        146 &        224 $\pm$         10 &       0.8 $\pm$       0.2 \\ 
DM Tau  &         18$^{+  34}_{-18}$ &        339$^{+  365}_{-339}$  &       -133 $\pm$         20 &        270 $\pm$         64 &    437 $\pm$         14 & 4.0 \\ 
RECX 11  &       1376 $\pm$          1 &        197 $\pm$          0 &  - & - & - & 2.0 \\
HD 135344B  &       2487 $\pm$       1109 &        346 $\pm$         52 &        118 $\pm$        151 &       2293 $\pm$       2168 &        158 $\pm$         95 & 0.8 $\pm$ 0.2\\ 
DE Tau  &       4835 $\pm$       4590 &        249 $\pm$         59 &        -13 $\pm$         78 &       1693$^{+  1914}_{-1693}$  &        225 $\pm$        152 & 1.1 $\pm$ 0.5 \\ 
 \hline
\end{tabular}
\\ \end{center}
\rule{0mm}{5mm}
Columns: (2) Peak intensity of rest gaussian emission component; (3) Width parameter for rest gaussian emission component; (4) Peak intensity of secondary gaussian emission component; (5) Width parameter for secondary gaussian emission component; (6) Local continuum for Ly$\alpha$
\end{sidewaystable}

\begin{sidewaystable}
{\bf \caption{\label{TBL:Lyman_parameters_2}Best-fit Ly$\alpha$ parameters (continued)}}
\begin{center}
\begin{tabular}{ccccc}\hline\\
 &  $v_{out}$ &  N$_{out}$ & N$_{ISM}$ & $\chi_\nu^2$ \\
Object  &  (km s$^{-1}$) &(10$^{19} cm^{-2}$) & (10$^{19} cm^{-2}$) &   \\ \hline
V4046 Sgr  &       -96 $\pm$         15 &       3.0 $\pm$       0.8 &       1.1 $\pm$       0.4 &        104 - 136 \\
DF Tau  &       -36 $\pm$          7 &      14.0 $\pm$       1.5 &      49.0 $\pm$       1.4 &       $<$ 1.08 \\
RECX 15  &       -85 $\pm$         36 &       1.1 $\pm$       1.0 &       0.9 $\pm$       0.2 &         2.2 - 20 \\
DM Tau  &       119 $\pm$        178 &       1.4 $\pm$       1.2 &      41.8 $\pm$       2.1 &         $<$ 1.16 \\
RECX 11  &       -99 $\pm$          4 &       1.9 $\pm$       0.1 &       3.3 $\pm$       0.1 &        $<$ 13 \\
HD 135344B  &       -99 $\pm$         26 &       5.6 $\pm$       2.7 &       7.1 $\pm$       1.6 &        0.76 - 7.1 \\
DE Tau  &      -103 $\pm$         56 &      11.4 $\pm$       7.7 &      34.8 $\pm$       8.3 &       $<$ 1.11 \\
 \hline
\end{tabular}\\
\end{center}
\rule{0mm}{5mm}
Columns: (2) Outflow absorber velocity; (3) Outflow absorber column density; (4) ISM absorber column density
\end{sidewaystable}

\begin{sidewaystable}
{\bf \caption{\label{TBL:CO_parameters}Best-fit CO parameters}}
\begin{center}
\begin{tabular}{cccccccc}\hline\\
 &  & Continuum  & $\Delta v_{broad}$  & log(N$_{CO}$) & T$_{CO}  $  & $v_{CO}$  & $\chi_\nu^2$\\ 
Object & Band & (FEFU) & (km s$^{-1}$) & (cm$^{-2}$) & (K) & (km s$^{-1}$) & \\\hline
V4046 Sgr & 14 - 3 &       3.1 &      27 $\pm$       1 &      18.55 $\pm$       0.45 &     241 $\pm$      41 &      -2.6 &  2.38 - 5.2 \\
V4046 Sgr & 14 - 4 &       3.5 &      30 $\pm$       3 &      18.47 $\pm$       0.41 &     295 $\pm$      87 &      -6.4 &  5.18 - 7.55 \\
DF Tau & 14 - 3 & 2.8 & 21 $\pm$ 1 &      18.56 $\pm$       .38 &     434 $\pm$     135 & 27 & $<$ 1.08 \\
RECX 15 & 14 - 3 &       0.4 &      16 $\pm$       1 &      18.50 $\pm$       0.53 &     318 $\pm$     131 &       1.0 & $<$ 1.14 \\
RECX 15 & 14 - 4 &       0.5 &      20 $\pm$       1 &      18.54 $\pm$       0.57 &     427 $\pm$     249 &       1.0 & 1.7 - 2.04 \\
DM Tau & 14 - 3 &       0.5 &      19 $\pm$       4 &      19.21 $\pm$       1.14&    833 $\pm$     562 &      35.0 & $<$ 1.18 \\
DM Tau & 14 - 4 &       0.8 &      17 $\pm$       3 &      19.08 $\pm$       0.77 &    775 $\pm$     445 &      26.2 & $<$ 1.14 \\
RECX 11 & 14 - 3 &       0.2 &      53 $\pm$      33 &      18.36 $\pm$       0.50 &     347 $\pm$     158 &      18.0 & $<$ 1.13 \\
HD 135344B & 14 - 3 &       1.0 &      41 $\pm$      14 &      18.58 $\pm$       0.93 &     148 $\pm$      66 &       5.0 & $<$ 1.13 \\
DE Tau & 14 - 3 &       0.3 &      26 $\pm$      20 &      18.35 $\pm$       1.03 &    793 $\pm$     696 &      30.0 & $<$ 1.14 \\
 \hline
\end{tabular}\\
\end{center}
\rule{0mm}{5mm}
Columns: (3) Local continuum for CO band; (4) FWHM of individual $J$ emission lines; (5) CO column density; (6) CO temperature; (7) CO LSR velocity
\end{sidewaystable}

\begin{table}
{\bf \caption{\label{TBL:1315_Sum} CO Band Summed Fluxes}}\begin{center}
\begin{tabular}{ccccc}\hline\\
Target &  $C_{(14-3)}$ &  Blue $F_{(14-3)}$ &  Red $F_{(14-3)}$ & $L(CO)$   \\
 & (FEFU) & (FEFU \AA)\tablenotemark{a} & (FEFU \AA) & (10$^{28}$ ergs s$^{-1}$)\\\hline
  & & & \\[-.1in]
V4046 &       3.30 $\pm$       0.68 &      47.36 $\pm$       0.39 &     20.33 $\pm$       0.28&     33.70 $\pm$       0.24\\
DFTAU &       3.73 $\pm$       1.02 &       4.69 $\pm$       0.31 &     17.38 $\pm$       0.40&     31.25 $\pm$       0.71\\
RECX15 &       0.54 $\pm$       0.31 &       2.11 $\pm$       0.12 &      3.35 $\pm$       0.15&      3.71 $\pm$       0.13\\
DMTAU &       2.20 $\pm$       1.02 &       9.19 $\pm$       0.41 &      4.21 $\pm$       0.39&     18.98 $\pm$       0.80\\
RECX11 &       0.33 $\pm$       0.24 &       2.18 $\pm$       0.11 &      1.01 $\pm$       0.10&      2.17 $\pm$       0.10\\
HD135344B &       1.28 $\pm$       0.73 &       6.38 $\pm$       0.39 &      3.18 $\pm$       0.35&     13.55 $\pm$       0.75\\
DETAU &       1.54 $\pm$       2.24 &       3.04 $\pm$       0.52 &      0.99 $\pm$       0.49&      5.70 $\pm$       1.01\\
GMAUR &       2.82 $\pm$       1.37 &       2.70 $\pm$       0.40 &      0.60 $\pm$       0.39&      4.68 $\pm$       0.79\\
HNTAU &       2.16 $\pm$       1.12 &       1.30 $\pm$       0.29 &      1.43 $\pm$       0.31&      3.87 $\pm$       0.60\\
DRTAU &      14.79 $\pm$       8.81 & 0 & 0 & 0\\
DKTAU &       4.14 $\pm$       4.90 &       4.57 $\pm$       1.49 &      8.16 $\pm$       1.74&     18.02 $\pm$       3.25\\
LkCa15 &       1.61 $\pm$       1.50 & 0 & 0 & 0 \\

 \hline
\end{tabular}
\tablenotetext{a}{1 FEFU \AA\ = 10$^{-15}$ ergs cm$^{-2}$ s$^{-1}$}
\end{center}
\end{table}

\end{document}